\documentclass[12pt]{article}
\usepackage{graphicx}
\begin{document}

\title{
 Chaplygin-Kalb-Ramond Quartessence}
\author{
Neven Bili\'c$^1$,
Gary B.\ Tupper$^2$,
and
Raoul D.\ Viollier$^2$
 \\
 $^1$Rudjer Bo\v{s}kovi\'{c} Institute, 10002 Zagreb, Croatia \\
E-mail: bilic@thphys.irb.hr \\
$^2$Department of Physics,
University of Cape Town, \\ Rondebosch 7701, South Africa\\
E-mail: parsons@physci.uct.ac.za
}
\maketitle

\begin{abstract}
Unified dark matter/energy models (quartessence) 
based upon the Chaplygin gas D-brane fail owing to the
 suppression of structure formation by the adiabatic speed of sound. Including string theory effects, in particular the Kalb-Ramond field which becomes massive via the brane,  we show how nonadiabatic perturbations allow successful structure formation.
\end{abstract}



\section{Introduction}
Cosmology today faces a mystery, dark matter, wrapped in an enigma: dark
energy. The existence of dark matter has been known for many years from
galactic rotation curves and cluster dynamics, with the lightest supersymmetric
particle being the leading suspect. Dark energy is a comparatively recent
discovery coming from observations of high-redshift supernovae \cite{perl1},
 and is
held responsible for the current accelerated Hubble expansion. Canonical
models of dark energy include a cosmological constant, quintessence, and
k-essence; for recent reviews, see \cite{peeb2}. Tight constraints come
from the WMAP \cite{benn3} observations of the cosmic microwave
background (CMB): $\Omega_{\rm DE}$ = 0.72,
$w_{\rm DE} = p_{\rm DE}/\rho_{\rm DE} < - $ 0.78,
$\Omega_{\rm DM}$ = 0.24, $w_{\rm DM} \simeq 0$,
with an impurity of 4\% of ordinary baryonic matter.

In contrast to the standard assumption that dark matter and dark energy
are distinct, there stands the hypothesis that both are different
manifestations of a single entity. The first definite model of this
type \cite{bil4} is based on the Chaplygin
gas, an exotic fluid with an equation of state
$p = - A/ \rho$,
whose cosmological possibilities were earlier noticed by Kamenshchik
 {\it et al.}\ \cite{kame5}.
Subsequently, the generalization to
$p = - A/ \rho^{\alpha}$, $0 \leq \alpha \leq 1$, was given
\cite{bent6} and the term `quartessence' coined \cite{mahl7}
to describe unified dark matter/energy models.

In a homogeneous model \cite{kame5}, the Chaplygin gas density is
$\rho (a) = \sqrt{A + B/a^{6}}$, where $B$ is an integration constant
and $a$ the scale factor normalized to unity today, thus interpolating
between dust
($\rho \sim a^{-3}$)
and a cosmological constant
($\rho \sim \sqrt{A}$).
The inhomogeneous Chaplygin gas   based on
a Zel'dovich type approximation has been proposed \cite{bil4}, and the picture has emerged that on caustics, where the density is high,
the fluid behaves as cold dark matter, whereas
in voids, $w=p/\rho$ is driven to the lower bound $-1$ producing
acceleration as dark energy.
 Soon, however, it has been shown that the simple Chaplygin
gas model does not work \cite{sand8,cart9,ame}. The physical reason is
that while the adiabatic speed of sound,
defined by
$c_{\rm s}^2 = \partial p / \partial \rho |_{S}$,
is small until $a \sim 1$, the accumulated comoving acoustic horizon
$d_{\rm s} = \int dt  c_{\rm s}/a \simeq H_{0}^{-1} a^{7/2}$
reaches MPc scales by redshifts of twenty, frustrating structure
formation even into the mildly nonlinear regime \cite{bil10}.
In the absence of caustic formation, the simple Chaplygin gas undergoes
damped oscillations that are in gross conflict with the observed mass
power spectrum \cite{sand8} and the CMB \cite{cart9,ame}. The problem is not
alleviated by the generalized Chaplygin gas,
$d_{\rm s} \simeq \sqrt{\alpha}
 \, H_{0}^{- 1}  a^{2 + 3 \alpha/2}$,
 unless
$\alpha$ is fine-tuned to unnaturally small values.

Such is the appeal of the quartessence idea that several variations
have been proposed to overcome the acoustic barrier. In particular,
Scherrer \cite{sche11} has explored a $k$-essence type model where the
Lagrangian has a local minimum as a function of the derivatives of the
$k$-essence field; such a model is equivalent to the ghost condensate
\cite{arka12} and hence shares its peculiarities \cite{krau13}.
Barreiro and Sen \cite{barr14} have suggested the generalized Chaplygin gas in a modified gravity approach, reminiscent of Cardassian models \cite{free15}. Yet, another deformation of the Chaplygin gas is described in \cite{nove16}.
One simple way to save the Chaplygin gas is to suppose that nonadiabatic  perturbations cause the
pressure perturbation $\delta p$ to vanish, and with it the acoustic horizon \cite{reis17}. To achieve this, it is necessary to add new  degrees of freedom
 which, to some extent, spoil 
 the simplicity of quartessence unification.

One of the most appealing aspects of the original Chaplygin gas model is that it is equivalent \cite{bil4}
to the Dirac-Born-Infeld description of a D-brane in string theory \cite{polc19}. In fact, string theory branes possess two features that are absent in the simple Chaplygin gas: (i) they support an Abelian gauge field $A_{\mu}$ reflecting
open strings with their ends stuck on the brane; (i) they couple to the (pull-back of) Kalb-Ramond \cite{kalb20}
antisymmetric tensor field $A_{\mu \nu} = - A_{\nu \mu}$ which, like the gravitational field $g_{\mu \nu}$, belongs to the closed string sector.
In this paper we  show that when these natural stringy features are incorporated into Chaplygin quartessence,
the basic scenario previously proposed in \cite{bil4} is realized 
owing to nonadiabatic perturbations.
It is important to stress that both the Dirac-Born-Infeld and Kalb-Ramond fields
originate from an ultimate theory in the context of string/M theory and,
unlike in quintessence models, each field affects both dark matter and dark
energy. Hence, dark matter and dark energy remain unified as in the simple
Chaplygin gas model.

We organize the paper as follows.
In Sec.\ \ref{nonadiabatic} we outline the basic ideas behind
the nonadiabatic perturbation mechanism.
In Sec.\ \ref{new} we describe the quartessence model based on
the Dirac-Born-Infeld Lagrangian extended by the Kalb-Ramond field.
Using  the Newtonian approach for simplicity, which is appropriate 
 when the pressure is small compared with the density, we demonstrate
 the cancellation of the pressure perturbation by the nonadiabatic
 perurbation mechanism. A discussion is given in Sec.\ \ref{discussion}
 and  Sec.\ \ref{conclusion} concludes the paper.

\section{Nonadiabatic perturbations}
\label{nonadiabatic}

It has been noted by Reis et al
\cite{reis17} that the root of the structure formation problem is the
term $\Delta \delta p$ in perturbation equations, equal to
$c_{s}^{2} \Delta (\delta \rho / \bar{\rho} )$ for adiabatic
perturbations, and if there are entropy perturbations such that
$\delta p = 0$, no difficulty arises.
This scenario, which is difficult to justify in the simple  Chaplygin gas model,
may be realized by introducing an extra degree of freedom, e.g.,
in terms of a quintessence-type scalar field $\phi$.

Suppose that the matter Lagrangian depends on two degrees of
freedom, e.g, a Born-Infeld scalar field $\theta$ and one additional
scalar field $\phi$. 
In this case, instead of a simple barotropic
form $p=p(\rho)$, the
equation of state involves the entropy density 
(entropy per particle) $s$ and may be  
written in the parametric form
\begin{equation}
p=p(\theta,\phi)\, ,
\label{eq201}
\end{equation}
\begin{equation}
\rho=\rho(\theta,\phi)\, ,
\label{eq202}
\end{equation}
\begin{equation}
s=s(\theta,\phi)\, .
\label{eq203}
\end{equation}
One of the  parameters, e.g., $\theta$ may be eliminated
in terms of $\rho$, so that the two remaining independent 
quantities,
the  pressure $p$ and the entropy density 
$s$, are expressed as functions of $\rho$ and $\phi$
and, hence, the corresponding perturbations are
\begin{equation}
\delta p=\delta \rho \frac{\partial p}{\partial \rho}
+\delta \phi \frac{\partial p}{\partial \phi}\, ,
\label{eq204}
\end{equation}
\begin{equation}
\delta s=\delta \rho \frac{\partial s}{\partial \rho}
+\delta \phi \frac{\partial s}{\partial \phi}\, .
\label{eq205}
\end{equation}
Then, the  speed of sound squared 
\begin{equation}
c_s^2\equiv \left. \frac{\delta p}{\delta \rho}\right|_{\delta s=0}=
\frac{\partial p}{\partial \rho}
- \frac{\partial s}{\partial \rho}
\left(\frac{\partial s}{\partial \phi}\right)^{-1}
\frac{\partial p}{\partial \phi}\, .
\label{eq206}
\end{equation}
is expressed as
the sum of two nonadiabatic terms: the usual 
$\partial p/\partial \rho$ and one additional term due to
the nonadiabatic  perturbation of the field $\phi$.
Thus, even for a nonzero $\partial p/\partial\rho$, the speed of sound
may  vanish 
if the second term on the right-hand side of 
(\ref{eq206})  cancels the first one.
This cancellation will take place if 
in the course of an adiabatic expansion, the perturbation 
$\delta \phi$ grows with $a$ in the same way as $\delta \rho$.
In this case, it is only a matter of adjusting initial conditions
of $\delta \phi$ with $\delta \rho$ to get $c_s=0$.

 To illustrate the  above general discussion,
 consider the 
 recently proposed 
 extended quart\-essence model \cite{bil18}
with the Lagrangian
\begin{equation}
{\cal{L}}_{\rm EQ} = \frac{1}{2}   g^{\mu \nu}   \phi_{, \mu}
\phi_{, \nu} - \sqrt{A}   {\rm e}^{-\omega \phi}
\sqrt{ 1 - g^{\mu \nu}   \theta_{, \mu}   \theta_{, \nu} }\, ,
\label{eq207}
\end{equation}
where
$\omega$ is a coupling constant
That is to say, we have replaced the constant brane
tension,
or the Chaplygin gas constant,
by the potential for a quintessence-type field $\phi$,
here taken to be exponential.
The model of Eq. (\ref{eq207}) is constructed so that
the perturbation of the pressure associated to the field $\theta$
is given by
\begin{equation}
\delta   p_{\theta} = - \bar{p}_{\theta}   \left(
\frac{\delta \rho_{\theta}}{\bar{\rho}_{\theta}}
 - 2\omega \delta \phi \right)\,
\label{eq301}
\end{equation}
and the above scenario is realized by $\delta \phi =
 \delta \rho_{\theta} /(2 \omega\bar{\rho}_{\theta})$
as an initial condition outside the causal horizon
$d_{c} = \int dt / a \simeq  H_{0}^{-1}  a^{1/2}$. 
However, the perturbation $\delta \phi$ satisfies 
\begin{equation}
\delta \ddot{\phi} + 3 H \delta \dot{\phi} -
\frac{1}{a^{2}}  \Delta \delta \phi \simeq 0 \,   ,
\label{eq001}
\end{equation}
with the solution in $k$-space
\begin{equation}
\delta \phi_k = a^{-3/4}  J_{3/2} (kd_{\rm c}) .
\label{eq302}
\end{equation}
Then, once the perturbations enter the causal horizon $d_{c}$ 
(but are still outside the
acoustic horizon $d_{s}$), $\delta \phi$ undergoes rapid damped
oscillations, so that the nonadiabatic perturbation associated with 
$\phi$ is destroyed. 
This means that the nonadiabatic
 perturbations are not automatically preserved except on long,  i.e.,
superhorizon, wavelengths where the simple Chaplygin gas has no problem anyway.
Thus,
 the extended quartessence model of Eq. (\ref{eq207})
 unfortunately
 does not solve the structure formation problem.
  
 In the following section we demonstrate how
  the Kalb-Ramond field
provides a mechanism for the desired cancellation of 
the nonadiabatic perturbations.
  
\section{New Quartessence}
\label{new}

We begin by recalling that the Kalb-Ramond field strength $H_{\mu \nu \alpha} = A_{\mu \nu, \alpha} + A_{\alpha \mu, \nu}
+ A_{\nu \alpha, \mu}$ is invariant under $A_{\mu \nu} \rightarrow A_{\mu \nu} + \xi_{\mu, \nu} - \xi_{\nu, \mu}$, with
a further $\xi_{\mu} \rightarrow \xi_{\mu} + \phi_{, \mu}$ gauge invariance.
 Thus, in four space-time dimensions there is one
 (6 - 4 - 1 = 1) degree of freedom,
  equivalent to a pseudoscalar.
   The D3 brane is described by
\begin{equation}
S_{\rm DBI}=\int d^4x\,\sqrt{- \det (g_{\mu \nu}) } \, {\cal{L}}_{\rm DBI} = - \sqrt{A} \:
\int d^4x\,\sqrt{-\det ( g_{\mu \nu}^{(\rm ind)} + B_{\mu \nu})  }\, ,
\label{eq002}
\end{equation}
where $g_{\mu \nu}^{(\rm ind)}$ is the induced metric on the brane,
$B_{\mu \nu} = A_{\mu \nu} + F_{\mu \nu}, F_{\mu \nu} = A_{\nu , \mu} - A_{\mu, \nu}$
(we have absorbed some factors into $A_{\mu}$).
The Dirac-Born-Infeld Lagrangian is
invariant under Kalb-Ramond gauge transformations if
$A_{\mu} \rightarrow A_{\mu} + \xi_{\mu}$ simultaneously.
Of course, the U(1) invariance
$A_{\mu} \rightarrow A_{\mu} + \varphi_{, \mu}$ gives
two degrees of freedom to $A_{\mu}$.
Through the Higgs mechanism, $A_{\mu \nu}$ may
eliminate $A_{\mu}$ to
yield three massive degrees of freedom,
as is made manifest by noting
that the Bianchi identity for
$F_{\mu \nu}$ gives
\begin{equation}
H_{\mu \nu \alpha} = B_{\mu \nu, \alpha} + B_{\alpha \mu, \nu} + B_{\nu \alpha, \mu} \, ,
\label{eq003}
\end{equation}
whereas in the temporal (unitary) gauge, $B_{0i} = 0$,
$B_{i j} = \epsilon_{i j k}  B^{k}$.
Similar observations have been made in \cite{chun21}.

The matter is then described by the Lagrangian
\begin{equation}
{\cal{L}} ={\cal{L}}_{H}+ {\cal{L}}_{\rm DBI}\, ,
\label{eq103}
\end{equation}
where
\begin{equation}
{\cal{L}}_{H} = \frac{1}{24 K} \, H_{\mu \nu \alpha}  H^{\mu \nu \alpha} ,
\label{eq004}
\end{equation}
with $K = 8 \pi G$ and the
normalization  fixed by string theory compactification.
Since the primary issue is structure formation, which takes
place before the onset of
the accelerated expansion, we evaluate ${\cal{L}}_{\rm DBI}$
using the Newtonian gravity approximation and the light-cone gauge
\cite{bor,jac}
for the D-brane embedding:
$X^{i} = x^{i}$,
$\left( X^{0} + X^{5} \right)  / \sqrt{2} = t$,
$\left( X^{0} - X^{5} \right)  / \sqrt{2} = \theta$.
Then $g_{00}^{(\rm ind)} = 2 (\Phi + \dot{\theta})$,
$g_{0i}^{(\rm ind)} = \theta_{, i}$ ,
$g_{ij}^{(\rm ind)} = - a^{2} \delta_{i j}$,
where $\Phi$ is the peculiar gravitational potential, and these
judicious choices yield
\begin{equation}
{\cal{L}} =
\frac{1}{4K}
\left[
\frac{\dot{B}^{k} \dot{B}^{k}}{a^{4}}
- \frac{ \left( B_{, k}^{k} \right)^{2} }{a^{6}}
\right]
             -  \sqrt{A \left(1 + \frac{B^k B^k}{a^4}\right) \left( 2 \dot{\theta} + 2\Phi  + \theta_{, i} \gamma^{i j}  \theta_{, j} \right)} \: ,
\end{equation}
with
\begin{equation}
\gamma^{i j} = \frac{a^{2} \delta^{i j} + B^{i} B^{j} / a^{2} }
                    {a^{4} + B^{k} B^{k} }
\end{equation}
being the symmetric part of the inverse of
$a^{2} \delta_{i j} - \epsilon_{i j k} B^{k}$.

The mass density is identified as
$\rho = - \partial {\cal{L}} / \partial \Phi$
which can be recast as a Bernoulli equation
\begin{equation}
\dot{\theta} + \Phi + \frac{1}{2} \theta_{, i} \gamma^{i j} \theta_{, j}
= \frac{ A ( 1 + B^{k} B^{k} / a^{4} ) }{2 \rho^{2}} \, .
\label{eq109}
\end{equation}
The peculiar gravitational potential is determined by Poisson's equation
\begin{equation}
\Phi_{, i i} = \frac{K}{2}  a^{2} ( \rho - \bar{\rho} ) \, ,
\label{eq009}
\end{equation}
and the field equation for $\theta$ yields
the conservation equation
\begin{equation}
\frac{1}{a^{3}} \frac{d}{dt} ( a^{3}  \rho ) +
( \rho \gamma^{i j} \theta_{, j} )_{, i} = 0 \, .
\label{eq010}
\end{equation}
Finally, the field equation for $B^{i}$ is
\begin{equation}
\frac{1}{a^3}
  \frac{d}{dt}
  \left( \frac{\dot{B}^{i}}{a} \right)
 -
\frac{
B_{,ji}^j}{a^{6}} + \frac{2 K A B^i}{a^{4}\rho}
            +  2 K \rho \;
\frac{
\theta_{, i}  B^{j}  \theta_{,j} / a^{2} - \gamma^{j k}
       \theta_{, j}  \theta_{, k} B^{i}}{
   a^{4} + B^{j} B^{j}} = 0 .
\label{eq011}
\end{equation}
If we make the decomposition
 $B^i=B^i_{\bot}+B^i_{\|}$ with the transverse part
satisfying $\partial_i B^i_{\bot}=0$,
the key point becomes evident: whereas the longitudinal part
  $B^{i}_{\|}$ suffers the
same problem as Eq.\ (\ref{eq001}),
the transverse part does not experience spatial gradients.

Next, let us see the implications for linear perturbation theory.
In addition to
$\delta = ( \rho - \bar{\rho} ) / \bar{\rho}$
and $\psi = \theta - \bar{\theta}$,
we count $B^{i} B^{j}$ as first order because
the Kalb-Ramond field, as a nonzero background, would be
incompatible with isotropy.
First, we show that 
once the perturbations enter the causal horizon,
the longitudinal part of $B^kB^k/a^4$ in 
Eq. (\ref{eq109}) undergoes damped oscillations,
similar to those experienced by $\delta \phi$ discussed in Sec.\
\ref{nonadiabatic}.
Retaining the dominant terms, Eq. (\ref{eq011})  becomes 
\begin{equation}
\frac{1}{a^3}\frac{d}{dt}\left(\frac{{\dot B}^i}{a}\right)-
\frac{B^j_{,ji}}{a^6}
+
\frac{2 K A}{\bar{\rho}} \: \frac{B^i}{a^{4}} = 0\,  .
\label{eq014}
\end{equation}
As $H^{2} = \left( \dot{a} / a \right)^{2} = K \bar{\rho} / 3$,
the last term in Eq.\ (\ref{eq014}) is of order
$H^{2} A/ \bar{\rho}^{2}$, so being
negligible compared with the first term which is ${\cal{O}} (H^{2})$,
until $a \sim 1$. Then, Eq. (\ref{eq014}) simplifies to
\begin{equation}
\frac{d}{dt}\left(\frac{{\dot B}^i}{a}\right)-
\frac{B^j_{,ji}}{a^3}=0,
\label{eq401}
\end{equation}
In the transverse components of this equation the last term is absent, 
 so the transverse solution reads
\begin{equation}
B_{\bot}^{i} ( a, \vec{x} ) \simeq c^{i} ( \vec{x} ) \int_{0}^{a} \frac{da}{H} = \frac{2}{5} \: \frac{c^{i} (\vec{x})}{H_{0} \Omega^{1/2} } \: a^{5/2} ,
\end{equation}
where $\Omega$ is the equivalent matter fraction at high redshift
and $c^{i} ( \vec{x} )$ are arbitrary functions of $\vec{x}$.
Thus,
\begin{equation}
\frac{B_{\bot}^{i} B_{\bot}^{i}}{a^4} = \frac{4}{25} \:
\frac{c^{i} c^{i}}{H_{0}^{2} \Omega} \: a
\label{eq402}
\end{equation}
grows linearly with the scale factor.

The longitudinal component of  Eq. (\ref{eq401}) in 
$k$-space reads
\begin{equation}
\frac{d}{dt}\left(\frac{{\dot B}_{\|}(k)}{a}\right)+
\frac{k^2}{a^3}B_{\|}(k)=0\, .
\label{eq406}
\end{equation}
with the solution
\begin{equation}
B_{\|}(k)=a^{5/4} J_{5/2}(kd_{\rm c}) ,
\label{eq110}
\end{equation}
Hence 
$B_{\|}$ oscillates inside $d_{\rm c}=2\Omega^{-1/2}
H_0^{-1}a^{1/2}$, with the amplitude 
increasing linearly with $a$.
As a consequence,
the longitudinal term $B_{\|}^2/a^4$  that enters 
the right-hand side of Eq. (\ref{eq109}) oscillates with an amplitude decreasing  
as $a^{-2}$ compared with 
the transverse term $B^i_{\bot}B^i_{\bot}/a^4$ that
grows linearly.

Since $  a^{3}\bar{\rho}$ is constant and $B_{\|}^2/a^4$ is damped inside $d_{\rm c}$, retaining only the transverse part in (\ref{eq109}), we obtain
\begin{equation}
\dot{\psi} + \Phi = \frac{A}{2 \bar{\rho}^{2}}
\left(\frac{B_{\bot}^{k} B_{\bot}^{k}}{ a^{4}} - 2 \delta \right),
\label{eq012}
\end{equation}
\begin{equation}
\dot{\delta} + \frac{1}{a^{2}} \, \psi_{, ii} = 0.
\label{eq013}
\end{equation}
Owing to Eq. (\ref{eq402})
  we can arrange
nonadiabatic
perturbations for overdensities only such that
$B_{\bot}^{k} B_{\bot}^{k} / a^{4} - 2 \delta = 0$
with the assurance that they will hold independent of scale
until $a \sim 1$. Combining Eqs.\ (\ref{eq009}), (\ref{eq012}), and
(\ref{eq013}) we find
\begin{equation}
\ddot{\delta} + 2 H \dot{\delta} - \frac{3}{2}  H^{2} \delta = 0 ;
\hspace{1cm} \delta > 0,
\end{equation}
which is our main result: the growing mode overdensities here do
not display the damped oscillations of the simple Chaplygin gas below
$d_{\rm s}$, but grow as dust. We remark that it matters little that this
applies only for $\delta > 0$ since the Zel'dovich approximation
implies that 92\% ends up in overdense regions. Still, it should be kept
in mind that underdensities will not behave as in \cite{reis17}.

\section{Discussion}
\label{discussion}
Clearly, there is an open question as to what inflation model can produce the
initial conditions in the Kalb-Ramond field and brane embedding to allow
subsequent structure formation. We believe this issue is likely to be
closely related
to another: namely, how does the Chaplygin-Kalb-Ramond model fit into the
braneworld picture? We emphasize that
$g_{\mu \nu}^{(\rm ind)} \neq g_{\mu \nu}$
describes a brane different  from that we inhabit, with the D3 Chaplygin
brane moving rapidly  until the onset of acceleration. Codimension-one
braneworlds have little space for that.

The light-cone gauge formalism used here can be taken as a starting point
for investigating questions beyond linear theory. In particular, one might
hope that the acoustic horizon does resurrect at very small scales to provide
the constant density cores seen in galaxies dominated by dark matter. Also,
the ``magnetic'' nature of the Kalb-Ramond field, reflected in the cometric
$\gamma^{i j}$, will generate rotation - it is pertinent to note that the
simple, nonrotating, self-gravitating Chaplygin gas has a scaling solution
$\rho (r)  \propto  r^{- 2/3}$, far different from the
$\rho (r)  \propto  r^{- 2}$ wanted for flat rotation curves.

Ultimately, the model must be confronted with large-scale structure and the CMB.
This necessitates a different gauge choice for $g_{\mu \nu}^{(\rm ind)}$.
For the simple Chaplygin gas where the standard model feels the metric $g_{\mu \nu}$,
it has been shown \cite{bil22} that a good fit to the data
is obtained if a vanishing sound speed
is imposed on the Zel'dovich fraction. New quartessence brings in an added feature:
the standard model can be considered as a non-Abelian Born-Infeld theory corresponding
to a stack of coincident D3 branes, identifying the Abelian factor with $A_{\mu}$.
Then the standard model experiences not the closed string metric $g_{\mu \nu}$, but
rather the open string metric \cite{seib23} $\gamma_{\mu \nu} = g_{\mu \nu} + B_{\mu \alpha}  B_{\nu}^{\alpha}$
constructed as the inverse of the symmetric part of $(g + B)^{-1}$.
This is reminiscent of the Einstein-Strauss
nonsymmetric gravity theory, albeit the governing action is different and so the
problems detailed in \cite{damo24} may well be avoided.

Finally, the Kalb-Ramond field alone does not provide for transient acceleration
as required by string theory. Rather than invoke an extra field as was done
in \cite{bil18}, one may note that there is little difference between the
Chaplygin gas D brane where $\sqrt{A}$ is constant, and tachyon models
describing unstable D branes where $\sqrt{A} \rightarrow V (\theta)$ provided
the potential is sufficiently flat \cite{cope25} while still providing
for eventual deceleration. The required flatness can be obtained through
warped
compactification\footnote{See, e.g., \cite{garo26}. Note that the limits obtained
there derive from the assumption that the photon field lives on the unstable
brane, contrary to the picture here.},
which also explains the small value $A^{1/8} \sim \mbox{MeV}$.
\section{Conclusions}
\label{conclusion}
In this paper we have shown how the Chaplygin gas model, when augmented by
string theory features, can realize the quartessence scenario proposed earlier
\cite{bil4}. No ad-hoc constructions
(such as generalized branes or generalized gravity) are required.
To conclude, we would echo the sentiments of \cite{reis17}: the day of
quartessence cosmology is not at an end, but rather at a new beginning.

\section*{Acknowledgments}
This
research is in part supported by the Foundation of Fundamental
Research (FFR) grant number PHY99-01241 and the Research Committee of
the University of Cape Town.
The work of N.B.\ is supported in part by
the Ministry of Science and Technology of the Republic of Croatia
under Contract No.\ 0098002.



\begin{thebibliography}{99}
%
\bibitem{perl1} Perlmutter {\it et al.},  Astrophys.\ J.\ {\bf 517}
(1999) 565; A.G.\ Riess {\it et al.},  Astron.\ J.\ {\bf 116} (1998) 1009.
%
\bibitem{peeb2} P.J.E.\ Peebles and B.\ Ratra,  Rev.\ Mod.\ Phys.\ {\bf 75} (2003) 599;
T.\ Padmanabhan,  Phys.\ Rept.\ {\bf 380} (2003) 235.
%
\bibitem{benn3} C.L.\ Bennett {\it et al.},  Astrophys.\ J.\ Suppl.\ {\bf 148} (2003) 1; D.N.\ Spergel {\it et al.},
ibid., 175.
%
\bibitem{bil4} N.\ Bili\'{c}, G.B.\ Tupper, and R.D.\ Viollier,
 Phys.\ Lett.\ B{\bf 535} (2002) 17.
%
\bibitem{kame5} A.\ Kamenshchik, U.\ Moschella, and V.\ Pasquier,  Phys.\ Lett.\ B{\bf 511} (2001) 265.
%
\bibitem{bent6} M.C.\ Bento, O.\ Bertolami, and A.A.\ Sen,
 Phys.\ Rev.\ D{\bf 66} (2002) 043507.
%
\bibitem{mahl7} M.\ Makler, S.Q.\ de Oliveira, and I.\ Waga,  Phys.\ Lett.\ B{\bf 555} (2003) 1.
%
\bibitem{sand8} H.B.\ Sandvik, M.\ Tegmark, M.\ Zaldarriaga, and I.\ Waga,  Phys.\ Rev.\ D{\bf 69} (2004) 123524.
%
\bibitem{cart9} D.\ Carturan and F.\ Finelli,  Phys.\ Rev.\ D{\bf 68} (2003) 103501;
R.\ Bean and O.\ Dor\'{e},  Phys.\ Rev.\ D{\bf 68} (2003) 023515.
\bibitem{ame}
 L.\ Amendola, F.\ Finelli, C.\ Burigana, and D.\ Carturan,
 JCAP {\bf 07} (2003) 005.
%
\bibitem{bil10} N.\ Bili\'{c}, R.J.\ Lindebaum, G.B.\ Tupper, and R.D.\ Viollier,  JCAP {\bf 11} (2004) 008.
%
\bibitem{sche11} R.J.\ Scherrer, Phys.\ Rev.\ Lett.\ {\bf 93} (2004) 011301.
%
\bibitem{arka12} N.\ Arkani-Hamed, H.C.\ Cheng, M.A.\ Luty, and S.\ Mukohyama,
 JHEP {\bf 05} (2004) 074.
%
\bibitem{krau13} A.\ Krause and S.-P.\ Ng, hep-th/0409241.
%
\bibitem{barr14} T.\ Barreiro and A.A.\ Sen, astro-ph/0408185.
%
\bibitem{free15} K.\ Freese and M.\ Lewis,  Phys.\ Lett. B{\bf 540} (2002) 1.
%
\bibitem{nove16} M.\ Novello, M.\ Makler, L.S.\ Werneck, and C.A.\ Romero, astro-ph/0501643.
%
\bibitem{reis17} R.R.R.\ Reis, I.\ Waga, M.O.\ Calv\~{a}o, and S.E.\ Jo\'{r}as,  Phys.\ Rev.\ D{\bf 68} (2003) 061302;
for an earlier related work see W.\ Hu, Astrophys.\ J.\ {\bf 506} (1998) 485.

%
\bibitem{polc19} J.\ Polchinski, String Theory, Cambridge University Press, Cambridge, 1998.
%
\bibitem{kalb20} M.\ Kalb and P.\ Ramond,  Phys.\ Rev.\ D{\bf 9} (1974) 2273.
%
\bibitem{bil18} N.\ Bili\'{c}, G.B.\ Tupper, and R.D.\ Viollier, astro-ph/0503428.
%
\bibitem{chun21} E.J.\ Chun, H.B.\ Kim, and Y.\ Kim; hep-ph/0502051.
%
\bibitem{bor}
M.\ Bordemann and J.\ Hoppe, Phys. Lett. B{\bf 325} (1994) 359;
N.\ Ogawa, Phys.\ Rev. D{\bf 62} (2000) 085023.
\bibitem{jac}
R. Jackiw, {\it Lectures on Fluid Mechanics} (Springer Verlag, Berlin, 2002).
 %
\bibitem{bil22} N. Bili\'{c}, R.J.\ Lindebaum, G.B.\ Tupper, and R.D.\ Viollier,
proceedings of the
Physical Cosmology Workshop, 15 Rencontres de Blois, France, 2003, to be published, astro-ph/0310181.
%
\bibitem{seib23} N.\ Seiberg and E.\ Witten,  JHEP {\bf 09} (1999) 032;
G.W. Gibbons and C.A.R. Herdeiro,  Phys. Rev.\ D{\bf 63} (2001) 064006.
%
\bibitem{damo24} T.\ Damour, S.\ Deser, and J.\ McCarthy,  Phys.\ Rev.\ D{\bf 47} (1993) 1541.
%
\bibitem{cope25} E.J.\ Copeland, M.R.\ Garousi, M.\ Sami, and S.\ Tsujikawa,  Phys.\ Rev.\ D{\bf 71} (2005) 043003.
%
\bibitem{garo26} M.\ Garousi, M.\ Sami, and S.\ Tsujikawa, hep-th/0412002.
\end{thebibliography}
\end{document}